\documentclass[aps,prl,twocolumn,showpacs,superscriptaddress]{revtex4-1}

\usepackage{dcolumn}   
\usepackage{bm}        
\usepackage{amssymb} 
\usepackage{graphicx}
\graphicspath{{./pict/}}

\newcommand\pictc[5]{\begin{figure}
                       \centerline{\vspace{-2mm}
                        \includegraphics[width=#1\columnwidth,height=0.7\textheight,keepaspectratio]{#3}}
                       \protect\caption{\protect\label{#4} #5}\vspace{-3mm}
                    \end{figure}            }

\newcommand\pict[4][1]{\pictc{#1}{!tb}{#2}{#3}{#4}}
\newcommand\rpict[1]{\ref{#1}}

\newcommand{\be}{\begin{equation}}
\newcommand{\ee}{\end{equation}}

\begin{document}

\title{Generation and near-field imaging of Airy surface plasmons}

\author{Alexander Minovich}
\affiliation{Nonlinear Physics Centre, Centre for Ultrahigh-bandwidth Devices for Optical Systems (CUDOS), Research School of Physics and Engineering, Australian National University, Canberra ACT 0200, Australia}

\author{Angela E. Klein}
\affiliation{Institute of Applied Physics, Friedrich-Schiller-Universit\"at Jena, Max-Wien-Platz 1, 07743 Jena, Germany}

\author{Norik Janunts}
\affiliation{Institute of Applied Physics, Friedrich-Schiller-Universit\"at Jena, Max-Wien-Platz 1, 07743 Jena, Germany}

\author{Thomas~Pertsch}
\affiliation{Institute of Applied Physics, Friedrich-Schiller-Universit\"at Jena, Max-Wien-Platz 1, 07743 Jena, Germany}

\author{Dragomir N. Neshev}
\affiliation{Nonlinear Physics Centre, Centre for Ultrahigh-bandwidth Devices for Optical Systems (CUDOS), Research School of Physics and Engineering, Australian National University, Canberra ACT 0200, Australia}

\author{Yuri S. Kivshar}
\affiliation{Nonlinear Physics Centre, Centre for Ultrahigh-bandwidth Devices for Optical Systems (CUDOS), Research School of Physics and Engineering, Australian National University, Canberra ACT 0200, Australia}
\date{\today}

\begin{abstract}
We demonstrate experimentally the generation and near-field imaging of nondiffracting surface waves--{\em plasmonic Airy beams}, propagating on the surface of a gold metal film. The Airy plasmons are excited by an engineered nanoscale phase grating, and demonstrate significant beam bending over their propagation. We show that the observed Airy plasmons exhibit self-healing properties, suggesting novel applications in plasmonic circuitry
and surface optical manipulation.
\end{abstract}

\maketitle



Airy wavepackets constitute a special class of nondiffracting waves that accelerate along parabolic trajectories and exhibit self-healing properties. First suggested more than 30 years ago in the pioneering work by Berry and Balazs~\cite{Berry:1979-264:AJP}, they became known to exist in various fields of physics~\cite{Vallee:2010}. Despite their well-known nondiffracting properties, Airy beams have been observed only recently in free-space optics~\cite{Siviloglou:2007-213901:PRL,Chong:2010-103:NatPhot} as one of several types of free-space nondiffracting waves that include Bessel and Mathieu beams. However, in low-dimensional systems such as graphene~\cite{Vakil:2011:arXiv} and magnetic films~\cite{Demidov:2010-984:NMat} the Airy beams are the only beams that do not spread with propagation. As such, Airy beams have recently been suggested theoretically in plasmonics~\cite{Salandrino:2010-2082:OL} and their manipulation through linear potentials~\cite{Liu:2011-1164:OL} suggests novel opportunities for
{\em nondiffractive plasmon optics}.

Surface plasmon polaritons (SPPs) are localized or propagating quasiparticles in which photons are coupled to the free electron oscillations in metals. The plasmon field is tightly confined to the metal surface decaying exponentially away from it. Such an effectively planar system with subwavelength confinement is very attractive for flatland photonics including the demonstration of optical nondiffracting SPPs. An analytical solution for the Airy plasmons in the paraxial approximation has recently been suggested~\cite{Salandrino:2010-2082:OL}, showing that Airy plasmons can be indeed possible despite the strong energy dissipation at the metal surface. While nonparaxial effects can play an important role in such purely evanescent Airy waves~\cite{Novitsky:2009-3430:OL}, the nonparaxiality could be neglected for practical applications~\cite{Salandrino:2010-2082:OL} such as energy routing through plasmonic boards, or directing nanoparticles on metal surfaces~\cite{Righini:2007-477:NatPhys}. As such the experimental demonstration of plasmonic Airy waves represents {\em an important milestone} in the field of plasmonics, however it still remains an unsolved challenge. Here,
for the first time to our knowledge, we demonstrate the direct generation of Airy nondiffracting surface plasmons and characterise their unique features, namely the propagation along self-accelerating trajectories and self-healing properties.

Practically, two major challenges for the experimental realization of Airy surface plasmons exist: (i) their {\em generation}, e.g. the appropriate coupling of free propagating light to surface plasmons with a complex field profile, and (ii) their {\em near-field observation}, such that the curved trajectory of the Airy SPPs can be observed before the plasmon decay with propagation. The generation problem is nontrivial. In experiments, only truncated Airy beams of a finite energy with exponentially attenuated profiles can be realized. In free-space, such beams are generated by imprinting of a cubic phase onto a Gaussian beam in the Fourier plane and subsequently transforming it by a lens back to the real space~\cite{Siviloglou:2007-213901:PRL, Chong:2010-103:NatPhot}. Unfortunately, this technique is not really applicable in the case of surface plasmons. While SPPs phase modulation and Fourier transformation can  be achieved through plasmonic lenses~\cite{Zentgraf:2011-151:NNano} and plasmon transformation optics~\cite{Huidobro:2010-1985:NL, Liu:2010-1991:NL}, the short propagation of SPPs would hinder the generation process. A more applicable way is to excite the Airy plasmons through direct transformation from free-space light into the Airy plasmons. Due to the wavevector mismatch between a SPP and a free-space wave, special coupling techniques are needed to perform such an operation. The commonly used methods, suitable for a broad beam excitation, could utilize either dielectric prisms or diffraction gratings. At a first glance, the diffraction pattern scheme seems to be quite a challenging way to obtain the required amplitude and phase distribution of the Airy SPPs~\cite{Salandrino:2010-2082:OL}. However, we demonstrate that direct grating coupling of plane-wave beams to Airy surface plasmons can be achieved by utilizing the uniqueness of the nondiffracting nature of Airy plasmons.

\pict[0.9]{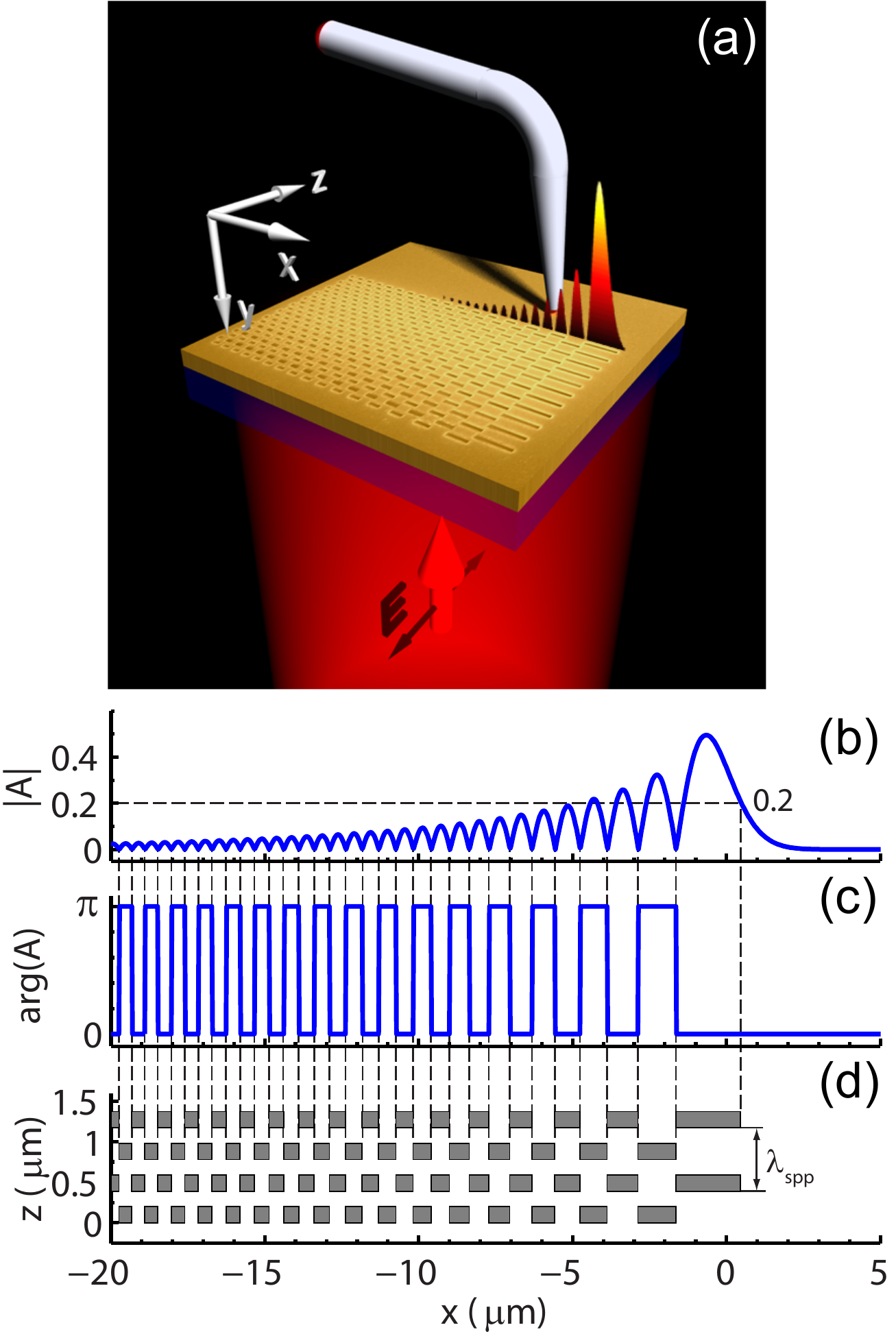}{fig1}{(color online) Generation of Airy plasmons: (a) Schematic of the experimental setup with the engineered grating.
The grating is composed of $11$ periods of $200$\,nm thick slits (in $z$-direction) and varying width in $x$-direction. The grating is excited from the glass substrate side with a broad Gaussian beam at $784$\,nm and polarization perpendicular to the slits. (b, c) Absolute value and phase of the amplitude function of the Airy plasmon. The main lobe half width is $x_0=700$\,nm.
(d) Grating geometry for generation of Airy plasmons. $\lambda_{\rm SPP}$ denotes the SPP wavelengths.}

In our approach for experimental generation of Airy plasmons we use a grating coupling scheme [Fig.~\rpict{fig1}(a)], with a grating pattern designed to imprint the phase profile of an Airy function, as shown in Figs.~\rpict{fig1}(b-d). No special measures to modulate the field amplitude have been taken, however we show that such modulation is not necessary due to the uniqueness of the Airy plasmons as the only nondiffracting beams on the surface. Most importantly, by using this excitation technique we rigorously demonstrate their distinct properties, namely nondiffracting nature, self-accelerating trajectory~\cite{Siviloglou:2007-213901:PRL}, and self-healing after passing through surface defects~\cite{Broky:2008-12880:OE}. We note that in the proposed technique the Airy beam is truncated by the size of the grating in $x$-direction. This truncation does not match the commonly used exponential apodization of the Airy wave~\cite{Siviloglou:2007-979:OL}, however the non-diffracting nature of Airy plasmon assists in the development of the correct field profile. Any additional radiation will quickly diffract due to the small, wavelength-scale size of the grating features used in our experiment. As such the non-diffracting Airy plasmon will dominate the beam propagation.

In our experiments we generate Airy plasmons on the air-gold interface of a $150$\,nm optically thick gold film deposited on a glass substrate [Fig.~\rpict{fig1}(a)] by a DC sputtering. The diffraction pattern is fbricated using a Focused Ion Beam (FIB FEI Helios 600) such that the metal was completely removed from the areas of the rectangular slits. The main lobe of the designed Airy plasmon has a half width of $x_0=700$\,nm. The envelope of the Airy packet represents an oscillating function with alternating positive maxima and negative minima of slowly decaying amplitude. The Airy function absolute value, therefore represents a sequence of peaks [Fig.~\rpict{fig1}(b)] that can be observed experimentally in the intensity distribution. The phase distribution of the Airy function thus shows alternating segments with values of $0$ and $\pi$ [Fig.~\rpict{fig1}(c)]. Thus, in the design of the grating, the segments determine the length of the pattern in the transverse direction $x$, and the required phase modulation is achieved by shifting of the next column of slits by a half of the SPPs wavelength [Fig.~\rpict{fig1}(d)]. To couple light from free-space into the SPP wave at normal incidence, the pattern is periodically repeated along $z$-direction with a period equal to the SPP wavelength. The grating is excited from the substrate side by a cw laser at $784$\,nm and power of $10$\,mW. The beam width onto the grating is $\sim200\,\mu$m, thus completely covering the grating pattern.

\pict[0.9]{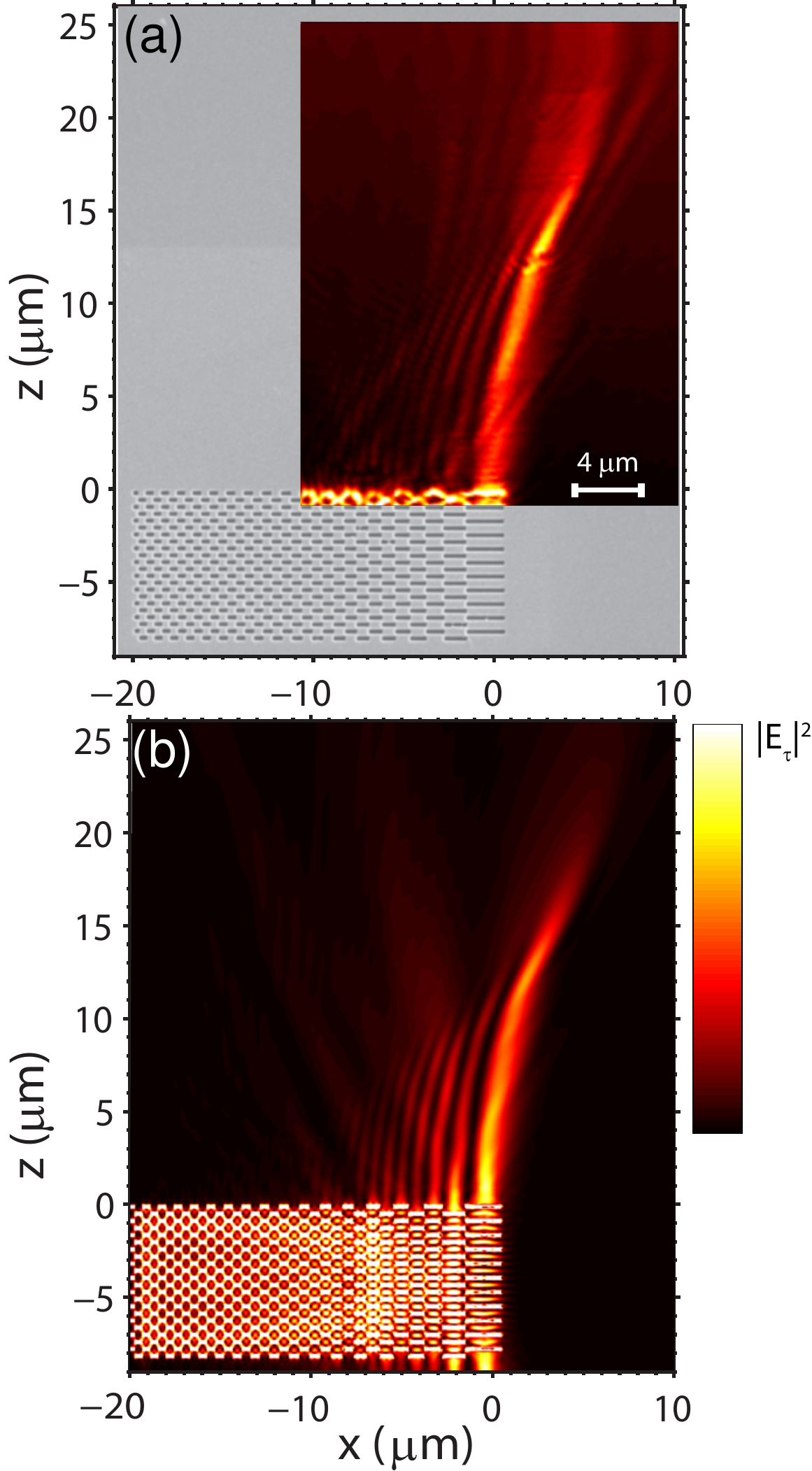}{fig2}{(color online) Near-field imaging of Airy plasmons. (a) Superimposed scanning electron microscope image of the grating pattern and the measured near-field intensity profile of the accelerating plasmon. The gold coated bent fiber probe used for the measurement is most sensitive to the tangential component $|E_\tau|^2$ of the electric field. (b) Numerically calculated distribution $|E_\tau|^2=|E_z|^2+|E_x|^2$ at the metal-air interface. The excitation wavelength is $\lambda_0=784$ nm.}

To test if the designed grating indeed couples the free propagating light into an Airy-type accelerating plasmon, next we perform near-field microscopy and image the electric field distribution at the metal surface. The near-field intensity distribution at the gold-air interface is characterized by a near-field scanning optical microscope (NSOM) (Nanonics MultiView 4000) operated in a contact mode. A bent gold-coated fiber tip with aperture size of $150$\,nm is used for light collection. The optical signal is then detected by the single photon counting module SPCM-AQR-14 (Perkin-Elmer). In Fig.~\rpict{fig2}(a) we superimpose a scanning electron microscope image of the fabricated grating with the measured NSOM image. The near-field distribution clearly shows that at the edge of the grating, we generate a surface plasmon that propagates on the metal surface along a curved trajectory for more than $20\,\mu$m. (In fact, two Airy plasmon beams are symmetrically generated on both sides of the grating.) Importantly, the main lobe bends by more than two lobe-widths before the wave looses its power due to the plasmon propagation losses. In Fig.~\rpict{fig2}(b) we compare the experimentally measured near-field profile with the one obtained theoretically by finite difference time-domain (FDTD) calculations and obtain excellent agreement between both. Even though both theory and experiment show some extra light that interferes with the generated Airy plasmon, this additional radiation diffracts quickly and its effect on the Airy plasmon is weak.
Numerical tests show that the amount of extra radiation depends most crucially on the size of the first widest slit of the grating and is largely minimized by truncating the slit at $0.2$ level of the Airy function [Fig.~\rpict{fig1}(b)].

\pict[0.99]{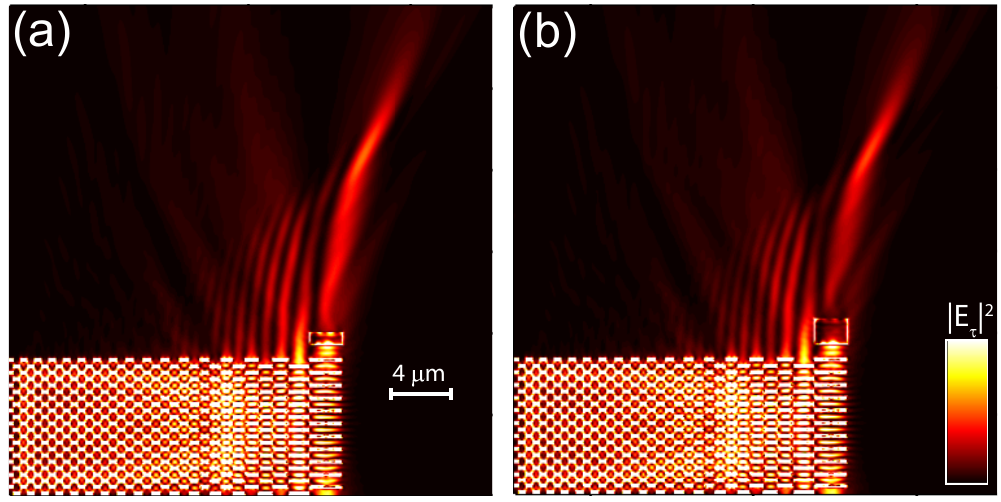}{fig5}{(color online) Airy plasmon self-healing: Numerically calculated near-field intensity profiles when a hole (shown with the dashed square) is made in the metal film at the path of the main Airy plasmon lobe. (a)~The hole length in $z$ direction is one SPP wavelength, (b)~the hole length is two SPP wavelengths. Excitation wavelength is $\lambda_0=784$\,nm. The Airy profile on a smooth metal surface is shown in Fig.~\rpict{fig2}(b).}

Importantly, in the experiments in Fig.~\rpict{fig2}(a) one can also observe the self-healing properties of the Airy plasmon. Namely that the Airy plasmon beam recovers quickly its shape after passing by a structural defect on the metal surface. Such defect is seen in Fig.~\rpict{fig2}(a) at the coordinate $(x,z) = (2.0,\,12.5)\,\mu$m. The measured field distribution shows that this surface defect disturbs the beam profile only locally, as expected for a nondiffracting beam. Further proofs on the strong self-healing properties of the Airy plasmons are provided by numerical simulations, where we artificially create a hole in the metal film in front of the main Airy lobe.
In Fig.~\rpict{fig5}(a,b) the Airy plasmon passes through a hole of one or two SPP wavelengths size, however in both cases it quickly recovers its shape shortly after the holes as compared to the Airy plasmon on a smooth metal surface [Fig.~\rpict{fig2}(b)]. These self-healing properties of the Airy plasmons make them specifically attractive for surface manipulation of nanoparticles, similar to the recent manipulation of micro-scale objects by free-space propagating Airy beams~\cite{Baumgartl:2008-675:NatPhot}.

The deflection of the main lobe extracted from the experimental and numerical data as well as the trajectory given by the paraxial approximation~\cite{Salandrino:2010-2082:OL}
are shown in Fig.~\rpict{fig3}(a). At the initial stage of propagation ($z\sim3\,\mu$m) we see some stronger variations of the beam deflection [Fig.~\rpict{fig3}(a)]. However after this initial stage, the main peak follows a smooth parabolic trajectory, which matches well the result of numerical calculations. This exactly parabolic trajectory is a direct signature of the nondiffracting nature of the generated plasmon beams.
Due to the non-paraxiality of the Airy plasmon ($\lambda_0=784$\,nm, $x_0=700$\,nm) both numerical and experimental results deviate from the analytical solution~\cite{Salandrino:2010-2082:OL} [Fig.~\rpict{fig3}(a), solid line].
Strictly speaking, the experimental field profile is not apodized exponentially as in Ref.~\cite{Salandrino:2010-2082:OL} and the effective apodization parameter $a$ can only be roughly estimated from the fit of the Airy tails.

The full width at half maximum (FWHM) of the main lobe is a good characteristic for the quality of an Airy plasmon and a verification of its non-diffracting nature. It has a constant value in the case of an ideal nondiffracting wave. In Fig.~\rpict{fig3}(b) we present the FWHM obtained from our experimental and numerical data. In both cases the beam width is affected by the extra radiation at the initial stage of propagation, however the main-lobe width [Fig.~\rpict{fig3}(b)] remains practically constant ($\sim 1.5\,\mu$m) over the diffraction-free zone of propagation. This zone is determined by the truncation of the Airy wavepacket~\cite{Siviloglou:2007-979:OL} due to the finite transverse size of our grating. In Fig.~\rpict{fig3}, the edge of the diffraction-free zone is marked with gray shading at around $15\,\mu$m. Outside this zone the main beam peak starts to quickly expand, loosing its Airy profile. Nevertheless, over the diffraction free Airy plasmon propagation we see a significant ($\sim3$ times) beam bending of the Airy plasmon trajectory.

While the measured and simulated FDTD results demonstrate a good agreement, the discrepancies between the analytical and experimental solutions can be attributed to the already mentioned difference in the beam apodization, as well as to the intrinsic for surface plasmons non-paraxial effects~\cite{Novitsky:2009-3430:OL}. Further improvement of the quality of the generated Airy plasmons can be possibly realized by the adjustment of the slit size in order to modulate the transmission through the grating.


\pict[0.9]{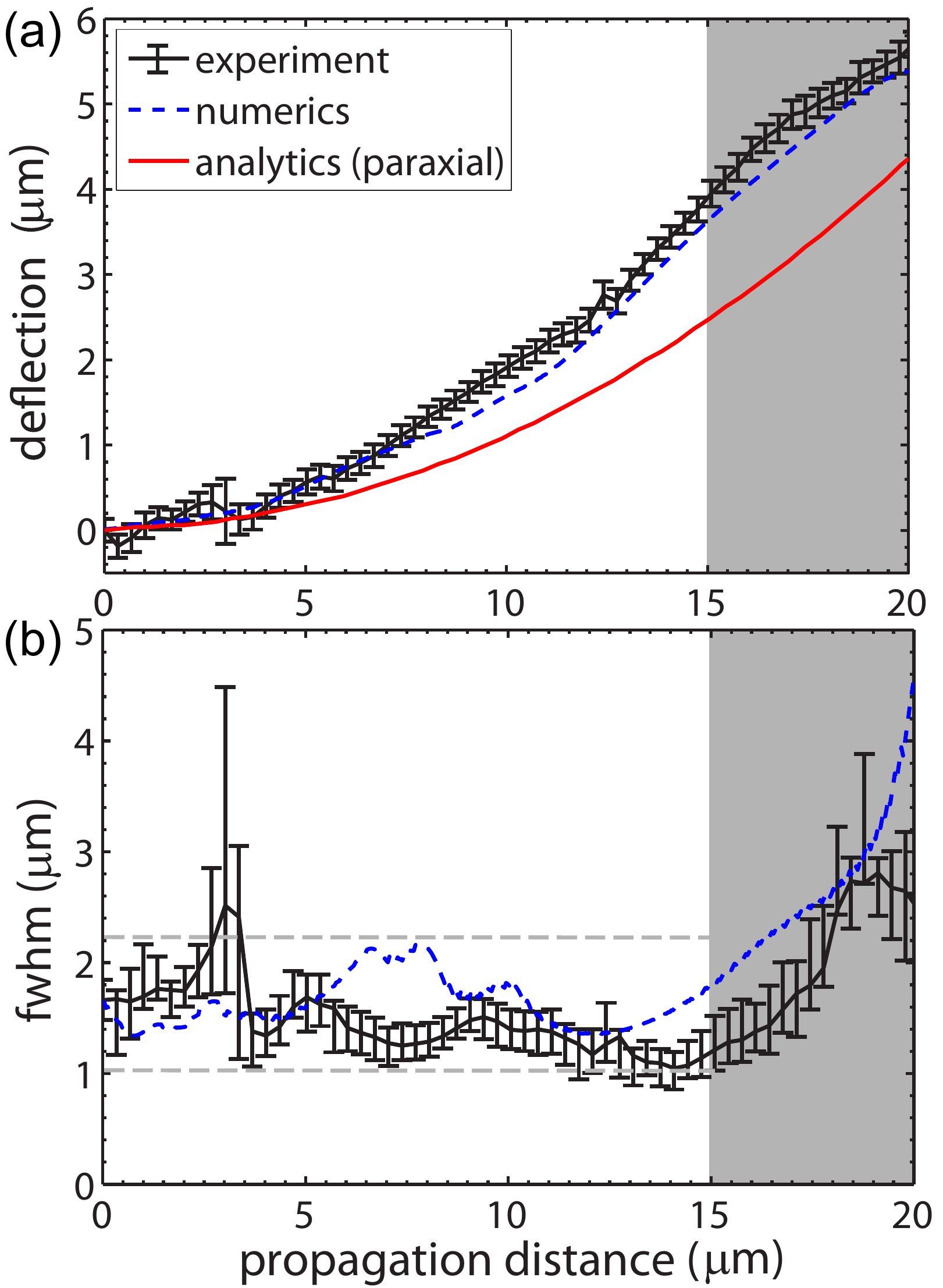}{fig3}{(color online) Airy plasmon trajectory: (a)~Main peak deflection versus propagation distance extracted from experimental data (solid curve with error bars), numerical data (dashed curve), and analytical solution for a paraxial Airy plasmon of $a=0.04$ (solid curve)~\cite{Salandrino:2010-2082:OL}.
(b)~Full width at half maximum of the main lobe versus propagation distance extracted from experimental data (solid curve with error bars) and numerical data (dashed curve).}

In conclusion, we have demonstrated the generation of nondiffracting Airy plasmons on a metal surface. We have employed specially designed diffraction grating that directly couples free-propagating light to plasmonic Airy beams. We have observed experimentally that the Airy plasmons propagate over parabolically curved trajectories and possess unique self-healing properties, being able to quickly recover after passing through obstacles and surface defects. We believe that the demonstration of Airy plasmon beams opens new opportunities for selective on-chip manipulation of nanoparticles~\cite{Righini:2007-477:NatPhys}, optical sensing, photonic circuitry~\cite{Ebbesen:2008-44:PhysTod}, and magneto-electronics. We anticipate that our results can open further opportunities for generation of subwavelength Airy beams in plasmonics or other surface systems, as recently attempted in spin waves~\cite{Schneider:2010-197203:PRL}. Finally, novel techniques for nonlinear Airy plasmon generation can be explored, combining the recent nonlinear generation of free-space Airy beams~\cite{Ellenbogen:2009-359:NatPhot} and nonlinear generation of surface plasmons~\cite{Simon:1974-1531:PRL, Palomba:2008-056802:PRL}.

We acknowledge support from the Australian National Computational Infrastructure, the ACT Node of the Australian Nanofabrication Facility, Go8-DAAD Joint Research Cooperation Scheme, and the Australian Research Council. A.K. acknowledges the support from the ``Jena School for Microbial Communication'' (JSMC).


\begin{thebibliography}{10}

\bibitem{Berry:1979-264:AJP}
M.~V. Berry and N.~L. Balazs, Am. J. Phys. \textbf{47}, 264 (1979).

\bibitem{Vallee:2010}
O.~Vall\'ee and M.~Soares, \emph{Airy Functions and Applications to Physics
  (2nd edition)} (Imperial College Press, 2010).

\bibitem{Siviloglou:2007-213901:PRL}
G.~A. Siviloglou, J.~Broky, A.~Dogariu, and D.~N. Christodoulides, Phys. Rev.
  Lett. \textbf{99}, 213901 (2007).

\bibitem{Chong:2010-103:NatPhot}
A.~Chong, W.~H. Renninger, D.~N. Christodoulides, and F.~W. Wise, Nat. Photon.
  \textbf{4}, 103 (2010).

\bibitem{Vakil:2011:arXiv}
N.~E. Ashkan~Vakil, arXiv:1101.3585v1  (2011).

\bibitem{Demidov:2010-984:NMat}
V.~E. Demidov, S.~Urazhdin, and S.~O. Demokritov, Nat Mater \textbf{9}, 984
  (2010).

\bibitem{Salandrino:2010-2082:OL}
A.~Salandrino and D.~N. Christodoulides, Opt. Lett. \textbf{35}, 2082 (2010).

\bibitem{Liu:2011-1164:OL}
W.~Liu, D.~N. Neshev, I.~V. Shadrivov, A.~E. Miroshnichenko, and Y.~S. Kivshar,
  Opt. Lett. \textbf{36}, 1164 (2011).

\bibitem{Novitsky:2009-3430:OL}
A.~V. Novitsky and D.~V. Novitsky, Opt. Lett. \textbf{34}, 3430 (2009).

\bibitem{Righini:2007-477:NatPhys}
M.~Righini, A.~S. Zelenina, C.~Girard, and R.~Quidant, Nat. Phys. \textbf{3},
  477 (2007).

\bibitem{Zentgraf:2011-151:NNano}
T.~Zentgraf, Y.~Liu, M.~H. Mikkelsen, J.~Valentine, and X.~Zhang, Nat Nano
  \textbf{6}, 151 (2011).

\bibitem{Huidobro:2010-1985:NL}
P.~A. Huidobro, M.~L. Nesterov, L.~Mart\'in-Moreno, and F.~J. Garc\'ia-Vidal,
  Nano Lett. \textbf{10}, 1985 (2010).

\bibitem{Liu:2010-1991:NL}
Y.~Liu, T.~Zentgraf, G.~Bartal, and X.~Zhang, Nano Lett. \textbf{10}, 1991
  (2010).

\bibitem{Broky:2008-12880:OE}
J.~Broky, G.~A. Siviloglou, A.~Dogariu, and D.~N. Christodoulides, Opt. Express
  \textbf{16}, 12880 (2008).

\bibitem{Siviloglou:2007-979:OL}
G.~A. Siviloglou and D.~N. Christodoulides, Opt. Lett. \textbf{32}, 979 (2007).

\bibitem{Baumgartl:2008-675:NatPhot}
J.~Baumgartl, M.~Mazilu, and K.~Dholakia, Nat. Photon. \textbf{2}, 675 (2008).

\bibitem{Ebbesen:2008-44:PhysTod}
T.~W. Ebbesen, C.~Genet, and S.~I. Bozhevolnyi, Phys. Today \textbf{61}, 44
  (2008).

\bibitem{Schneider:2010-197203:PRL}
T.~Schneider, A.~A. Serga, A.~V. Chumak, C.~W. Sandweg, S.~Trudel, S.~Wolff,
  M.~P. Kostylev, V.~S. Tiberkevich, A.~N. Slavin, and B.~Hillebrands, Phys.
  Rev. Lett. \textbf{104}, 197203 (2010).

\bibitem{Ellenbogen:2009-359:NatPhot}
T.~Ellenbogen, N.~Voloch-Bloch, A.~Ganany-Padowicz, and A.~Arie, Nat. Photon.
  \textbf{3}, 395 (2009).

\bibitem{Simon:1974-1531:PRL}
H.~J. Simon, D.~E. Mitchell, and J.~G. Watson, Phys. Rev. Lett. \textbf{33},
  1531 (1974).

\bibitem{Palomba:2008-056802:PRL}
S.~Palomba and L.~Novotny, Phys. Rev. Lett. \textbf{101}, 056802 (2008).

\end{thebibliography}

\end{document}